\documentclass[%
 reprint,
 amsmath,amssymb,
 aps,
 prl,
]{revtex4-1}

\usepackage{graphicx}
\usepackage{dcolumn}
\usepackage{bm}
\usepackage{natbib}

\newcommand{\confestimate}{\widehat{\{X\}}}

\newcommand{\ith}{$i^{\text{th}}$}
\newcommand{\poi}{$p\left( O_i \right)$}
\newcommand{\setpoi}{$\{ p\left( O_i \right)\}$}
\newcommand{\setX}{\{X\}}
\newcommand{\oi}{o^{(i)}}
\newcommand{\jointconditional}{p\left(\setX|O_1, O_2,\ldots,O_N\right)}

\begin{document}

\title{Inference of joint conformational distributions from separately-acquired experimental measurements}

\author{Jennifer M. Hays}
\affiliation{
 Departments of Biomedical Engineering and Molecular Physiology, University of Virginia}
\author{Emily Boland}
\affiliation{Departments of Biomedical Engineering and Molecular Physiology, University of Virginia}
\author{Peter M. Kasson}
\affiliation{%
 Departments of Biomedical Engineering and Molecular Physiology, University of Virginia}
\affiliation{
 Science for Life Laboratory, Department of Cell and Molecular Biology, Uppsala University
}

\date{\today}

\begin{abstract}
Many biomolecules have flexible structures, requiring distributional estimates of their conformations. Experiments to acquire distributional data typically measure pairs of labels separately, losing information on the joint distribution. These data are assumed independent when estimating the conformational ensemble. We developed a method to estimate the true joint distribution from separately acquired measurements, testing it on two biological systems. This method accurately reproduces the joint distribution where known and generates testable predictions about complex conformational ensembles.
\end{abstract}

\maketitle


Flexible protein structures are often determined using multiple spectroscopic measurements that are performed independently on different parts of the protein but considered jointly\cite{hays_refinement_2018, hays_hybrid_2019, marinelli_ensemble-biased_2015, roux_restrained-ensemble_2013}. These independent measurements are treated as uncorrelated, while in actuality there likely exists a correlation structure in conformation probability space. Here, we introduce a means to estimate the joint distributions including any correlation structure.  Such estimation is enabled by a stochastic resampling approach that estimates the free energy of restraining the protein to any individual conformation using non-equilibrium work measurements in simulations.

Flexible proteins play a critical role in a wide variety of cellular processes, including flexible recognition events during infection and in signal transduction pathways. This flexibility is essential to biological function\cite{motlagh_ensemble_2014, norman_steroid-hormone_2004,wright_intrinsically_2015, dyson_intrinsically_2005}. Experimental methods that have traditionally been used to study the structural ensembles of biological systems, like X-ray crystallography and nuclear magnetic resonance (NMR) spectroscopy, tend to reduce the ensemble to just a few low energy states in order to achieve high-resolution structures\cite{boehr_role_2009,jimenez_protein_2004,willcox_tcr_1999}. As awareness has increased of the fundamental role structural heterogeneity plays in biological function, new experimental methods have been developed to report directly on full ensembles. Methods such as double electron-electron resonance (DEER) spectroscopy and single molecule F{\"o}rster resonance energy transfer (smFRET) provide distance distributions between labeled amino acids, and thus yield quantitative information on conformational populations in a sample\cite{bernado_structural_2007,wei_protein_2016,levin_ensemble_2007,bonvin_conformational_1995}. However, these methods come with an important set of challenges, described below.

Label-based experiments that yield distributional data are severely restricted in the number of labels that can be measured simultaneously, leading to two major limitations: since each distribution requires a separate, time-consuming experiment, the data tend to be sparse over atomic coordinates, and separately measured label sets do not provide information on the joint distribution. Recent efforts have ameliorated the former limitation by optimizing label placement to ensure maximally informative measurements\cite{mittal_predicting_2017,jeschke_characterization_2012,hays_refinement_2018}, but little progress has been made in handling the latter. Here we present a general method for inferring joint probability distributions from separately-acquired measurements. The method not only estimates the correlation structure of the experimental distributions, but also provides a direct way to infer the conformational ensemble of interest.

We first lay out the theoretical basis for the approach, then apply the method to two example systems: a toy model of an alternating-access transporter and the soluble N-ethylmaleimide-sensitive factor attachment receptor (SNARE) protein syntaxin-1a. In the case of the alternating transporter, where the joint distribution is known, we find that our method accurately reproduces the joint distribution and correctly estimates the true conformational ensemble. In the case of syntaxin, DEER data have been acquired, but the joint distribution is unknown. We find that EESM converges stably to a final estimate of the joint distribution which differs significantly from the convolution of the experimental distributions. The new joint distribution provides novel, testable structural data on the syntaxin that may be used to guide future experiments. Although we have chosen to demonstrate the approach using specific biological systems, the method will estimate joint probability distributions and conformational ensembles of any system for which distributional data can be obtained.

Let us denote a set of separately measured probability distribution functions \setpoi, where $O_i$ is a random variable representing the observable of interest. In this convention, particular values of $O_i$ are denoted $\oi_j$. In the applications presented later, each \poi ~is a single DEER distribution and $O_i$ is the distance variable of the \ith ~pair of atoms. We wish to estimate not only the joint probability distribution $p(O_1, O_2,\ldots,O_N)$, but the conformational ensemble $\setX$ which optimally reproduces the joint distribution. This inference problem can be stated in terms of conditional probabilities: what is the probability of an ensemble $\setX$ given a set of distance variables, i.e., what is $\jointconditional$? The joint probability distribution is proportional to the free energy difference of the desired ensemble from some (arbitrary) reference ensemble:
\begin{equation}
    \jointconditional\propto e^{-\beta\Delta G\left(\setX|O_1, O_2,\ldots,O_N\right)}
\end{equation}
If each random variable $O_i$ can take on values $\left\{\oi_j\right\}$ with probability $p\left(\oi_j\right)$, then the probability of observing a particular conformation given a specific set of distances $\left\{o^{(1)}_{j=k},...,o^{(N)}_{j=m}\right\}$ is trivially:
\begin{equation}
p\left(x|o^{(1)}_{j=k},...,o^{(N)}_{j=m}\right) \propto e^{-\beta \Delta G\left(x|o^{(1)}_{j=k},...,o^{(N)}_{j=m}\right)}
\end{equation}

The challenge then lies in determining the free energy landscape $\Delta G$ as a function of the experimental observables. In some cases, it may be possibly to calculate this free energy analytically or via thermodynamic integration, but in general, it is prohibitively expensive to directly calculate the equilibrium free energy because of the large number of degrees of freedom and the slow relaxation timescales involved. Instead, the most robust and general method for calculating this free energy is via non-equilibrium sampling and use of the Jarzynski equality: $e^{-\beta\Delta G}=\langle e^{-\beta W}\rangle$\cite{jarzynski_nonequilibrium_1997}. We detail below how to leverage the Jarzynski equality and the experimental data to estimate the free energy landscape.

We previously developed a methodology, bias-resampling ensemble refinement (BRER), to incorporate distributional data into molecular dynamics (MD) simulation\cite{hays_hybrid_2019}. The original method assumes that all \setpoi ~are independent, but a simple extension of this formalism enables estimation of the joint distribution. The original BRER method is an iterative approach as follows: 

\begin{enumerate}
    \item randomly sample a conformation $x$ from the current ensemble estimate $\confestimate$
    \item select a set of observables, $\left\{O_1=o^{(1)}_{j=k},\ldots,O_N=o^{(N)}_{j=m}\right\}$, via probability-weighted draws from the experimental distributions \setpoi.
    \item run a biased MD simulation to constrain the conformation $x$ such that all $O_i=\oi_j$
    \item update the estimate $\confestimate$ with the final conformation $x$
\end{enumerate}
The method is trivially parallelized by drawing multiple conformations $\{x\}$ in a single iteration and applying the constraints to each $x$. 

To estimate the free energy of a set of conformations $\{x\}$ given a set of observables $\left\{\oi_j\right\}$, we can leverage the data from the biased MD runs of step (3). Because we use a simple linear biasing potential, it is trivial to calculate the work done on the ensemble to enforce the constraints. We can thus apply this simple linear bias to restrain the system to a single point in observable space, then use the nonequilibrium work to estimate equilibrium free energy differences via Jarzynski's equality. Specifically:
\begin{equation}
e^{-\beta\Delta G\left(x|o^{(1)}_{j=k},\ldots,o^{(N)}_{j=m}\right)}
 = \left\langle e^{-\beta W\left(x'\rightarrow x | o^{(1)}_{j=k},\ldots,o^{(N)}_{j=m}\right)}\right\rangle_{x'\in \{X'\}} 
\end{equation}
where $\{X'\}$ is a reference ensemble, in this case the estimated equilibrium ensemble assuming uncorrelated observables.

The general method for calculating both the joint distribution and the conformational ensemble from simulation, which we call Ensemble Estimation from Separate Measurements (EESM), can be summarized as follows:

\begin{enumerate}
    \item Draw a set of conformations $\left\{x'\right\}$ from a reference ensemble.
    \item Select a set of specific observable values $\left\{o^{(i)}_j\right\}$ via stochastic draws from each \poi.
    \item Apply a linear biasing potential such that $O_i = o^{(i)}_j$ for all $x'$.
    \item Calculate the work done in (3) and, consequently, the  probabilities $p\left(x|o^{(1)}_{j=k},\ldots,o^{(N)}_{j=m}\right)$.
    \item Repeat 1-4 until the distribution $\jointconditional$ has been estimated.
\end{enumerate}
This method is demonstrated below for a simple biological model system.

\begin{figure}
    \centering
    \includegraphics[width=0.45\textwidth]{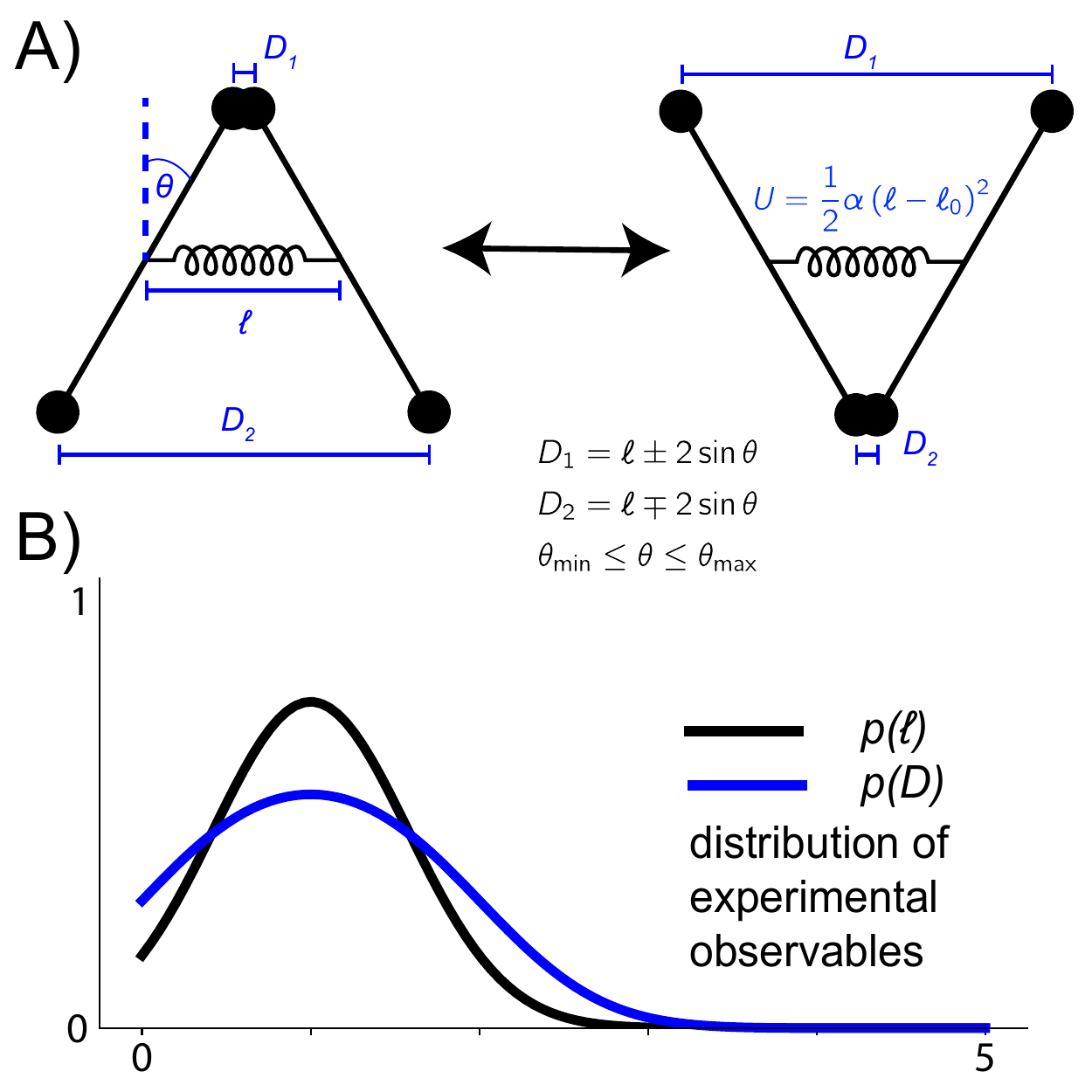}
    \caption{\textbf{Toy model of an alternating-access transporter used to test EESM.} A simplified model of a ``flexible" alternating-access transporter is schematized in (A). Experimental measurements of $D_1$, $D_2$, and $\ell$ would yield the distributions shown in (B).}
    \label{fig:corr1}
\end{figure}

Alternating-access transporters are a class of membrane proteins that transport their substrates by switching between outward-facing and inward-facing conformations\cite{widdas_inability_1952,jardetzky_simple_1966,rees_abc_2009,abramson_lactose_2003,ward_flexibility_2007,wilkens_structure_2015}. In order to test the EESM approach, we studied a simple model of a “flexible” alternating transporter (Fig \ref{fig:corr1}A). The model consists of two rigid rods connected at their midpoints by a spring with constant $\alpha$. The rods rotate about their midpoints subject to two constraints: they mirror each other’s rotation (the ``channel" of the transporter is a symmetry axis) and the angle of rotation $\theta$ is constrained to a range $[\theta_{\min}, \theta_{\max}]$. For a given channel width $\ell$, all permitted values of $\theta$ have equal energy, while those outside the permitted range have infinite energy.

\begin{figure}
    \centering
    \includegraphics[width=0.45\textwidth]{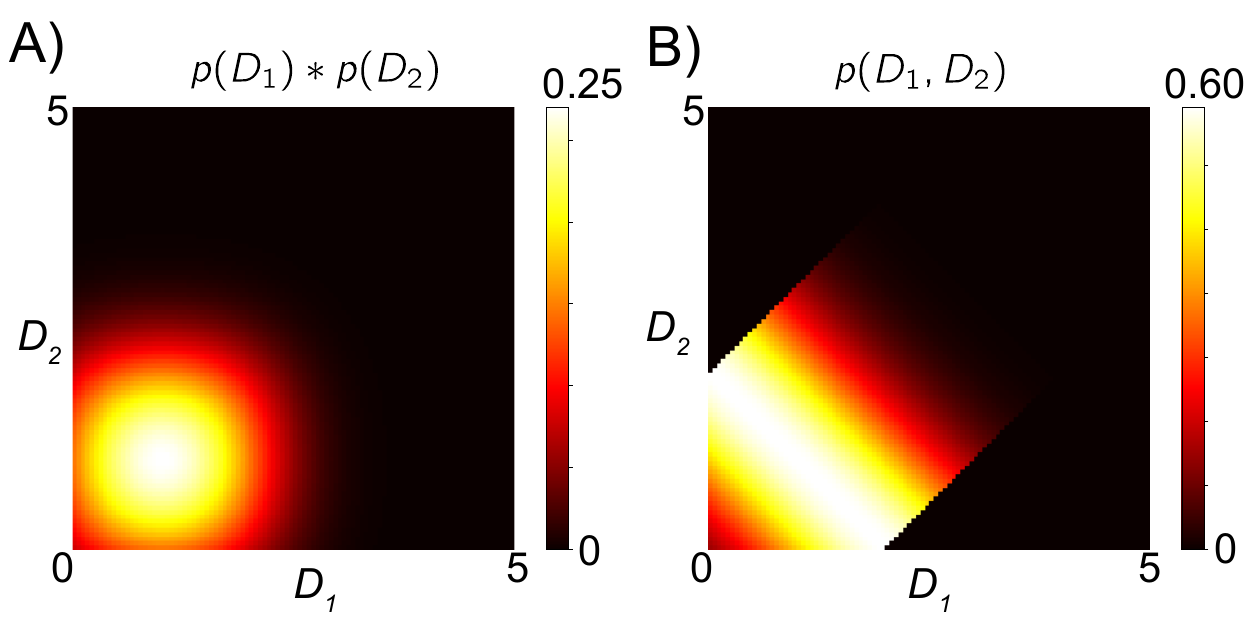}
    \caption{\textbf{Variables $D_1$ and $D_2$ are correlated, and this correlation is critical to estimating the joint distribution}. Plotted in (A) is the joint distribution if these two distance distributions from Fig \ref{fig:corr1} were uncorrelated and in (B) is the true joint distribution.  The assumption that $D_1$ and $D_2$ are independent variables leads to an incorrect estimate of their joint distribution.}
    \label{fig:corr2}
\end{figure}

We can imagine performing three separate experiments on the transporter to try to estimate its conformational ensemble: one that measures the width of the channel midpoint $\ell$, one that measures the distribution of the ``inward-facing" mouth of the channel ($D_1$ of Fig \ref{fig:corr1}A), and one that measures the ``outward-facing" mouth of the channel ($D_2$ of Fig \ref{fig:corr1}A). The results of these hypothetical experiments are shown in Fig \ref{fig:corr1}B. Without any additional information, we would assume that the separately measured variables $D_1$ and $D_1$ are independent and we would estimate the joint probability distribution as shown in Fig \ref{fig:corr2}A. However, because of the constraints imposed on the channel, the true joint distribution is dramatically different (Fig \ref{fig:corr2}B).

\begin{figure}
    \centering
    \includegraphics[width=0.45\textwidth]{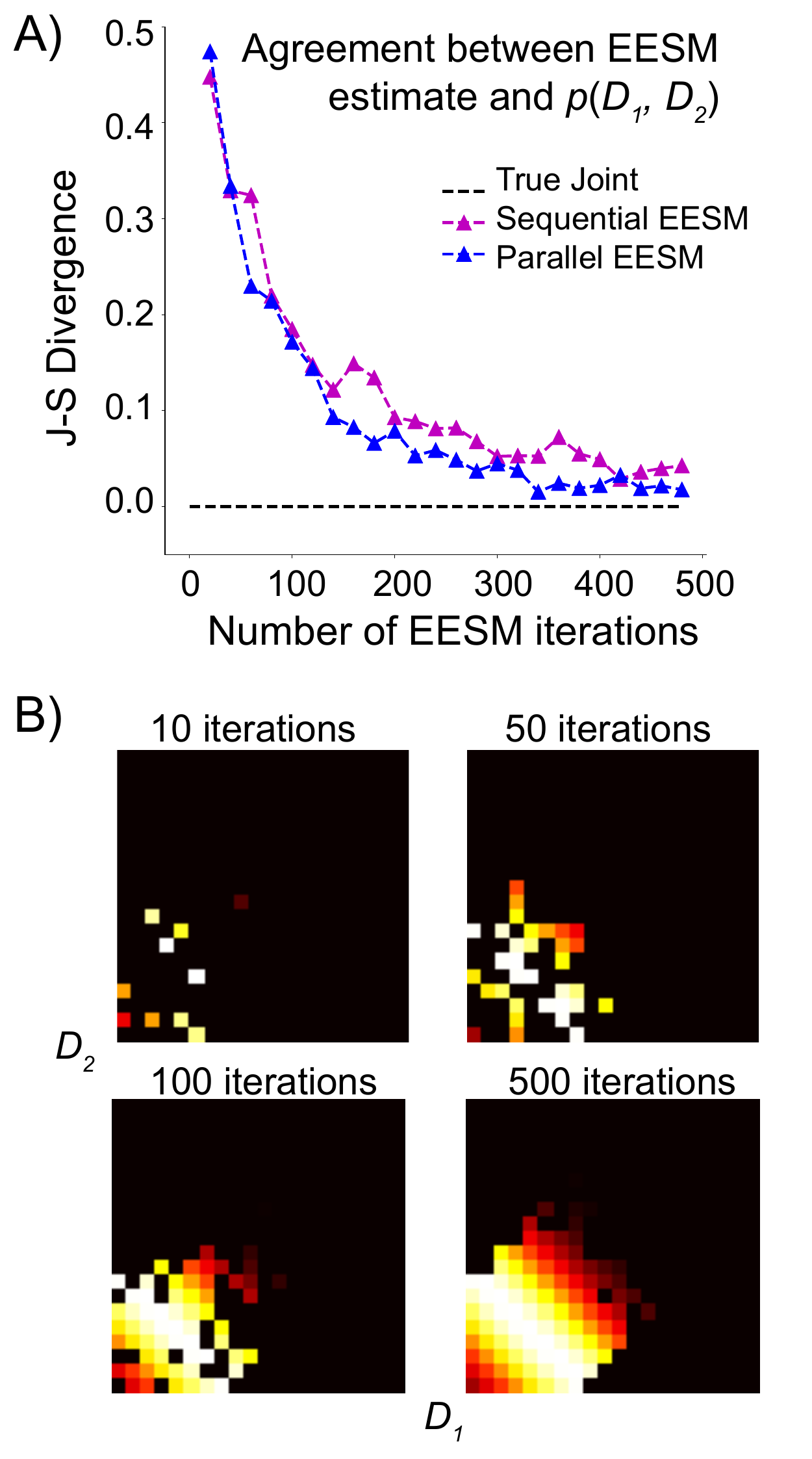}
    \caption{\textbf{EESM accurately reproduces the joint probability distribution for a simple alternating-access transporter.} Using EESM, we build a stochastic estimate of the true joint distribution $P(D_1,D_2)$ over 500 iterations of sampling performed sequentially and in parallel. Agreement between the estimate and true distribution is quantified as Jensen-Shannon divergence in (A); examples of the estimates over multiple iteration numbers are shown in (B).}
    \label{fig:corr3}
\end{figure}

In order to estimate the true distribution from only the experimental observables, we performed 500 aggregate iterations of EESM (details are provided in the Supplemental Materials). As the number of iterations increases, the estimate of the joint distribution approaches the true distribution. This is quantified via Jensen-Shannon divergence in Fig \ref{fig:corr3}A and illustrated as plots of the joint distribution in Fig \ref{fig:corr3}B. This simple but powerful example demonstrates that the method can indeed recover the correlation structure of separately measured distributions.

Refinement of the SNARE protein syntaxin-1a presents a significantly more challenging problem because it requires knowledge of many more degrees of freedom. SNARE proteins drive neuronal vesicle fusion and thus synaptic neurotransmission\cite{jahn_snares_2006,carr_taming_2001,misura_three-dimensional_2000,chen_snare-mediated_2001,gerber_conformational_2008}. Syntaxin exhibits a complex open/closed conformational equilibrium believed to regulate SNARE complex assembly: “open” syntaxin is able to form SNARE complexes, but “closed” is not\cite{lerman_structural_2000,margittai_single-molecule_2003}. The closed state has been characterized experimentally\cite{chen_nmr_2008}, but the open state ensemble remains underdetermined\cite{rizo_snares_2002,liang_prefusion_2013,wang_conformational_2017}. Thus, refining the open state conformational ensemble would provide insight into the mechanism of SNARE complex assembly and the regulation of neurotransmission. 

We estimated the conformational ensemble of syntaxin using EESM and a set of three published DEER measurements. To obtain pair-wise distance distributions, DEER measurements must be obtained in a sequential fashion, meaning that information on the joint distribution is lost. A previously previously published estimate of the syntaxin ensemble assumes that the DEER distributions are independent\cite{hays_hybrid_2019}; here, we refine the estimate further by explicitly calculating the joint distribution. 

\begin{figure}
    \centering
    \includegraphics[width=0.45\textwidth]{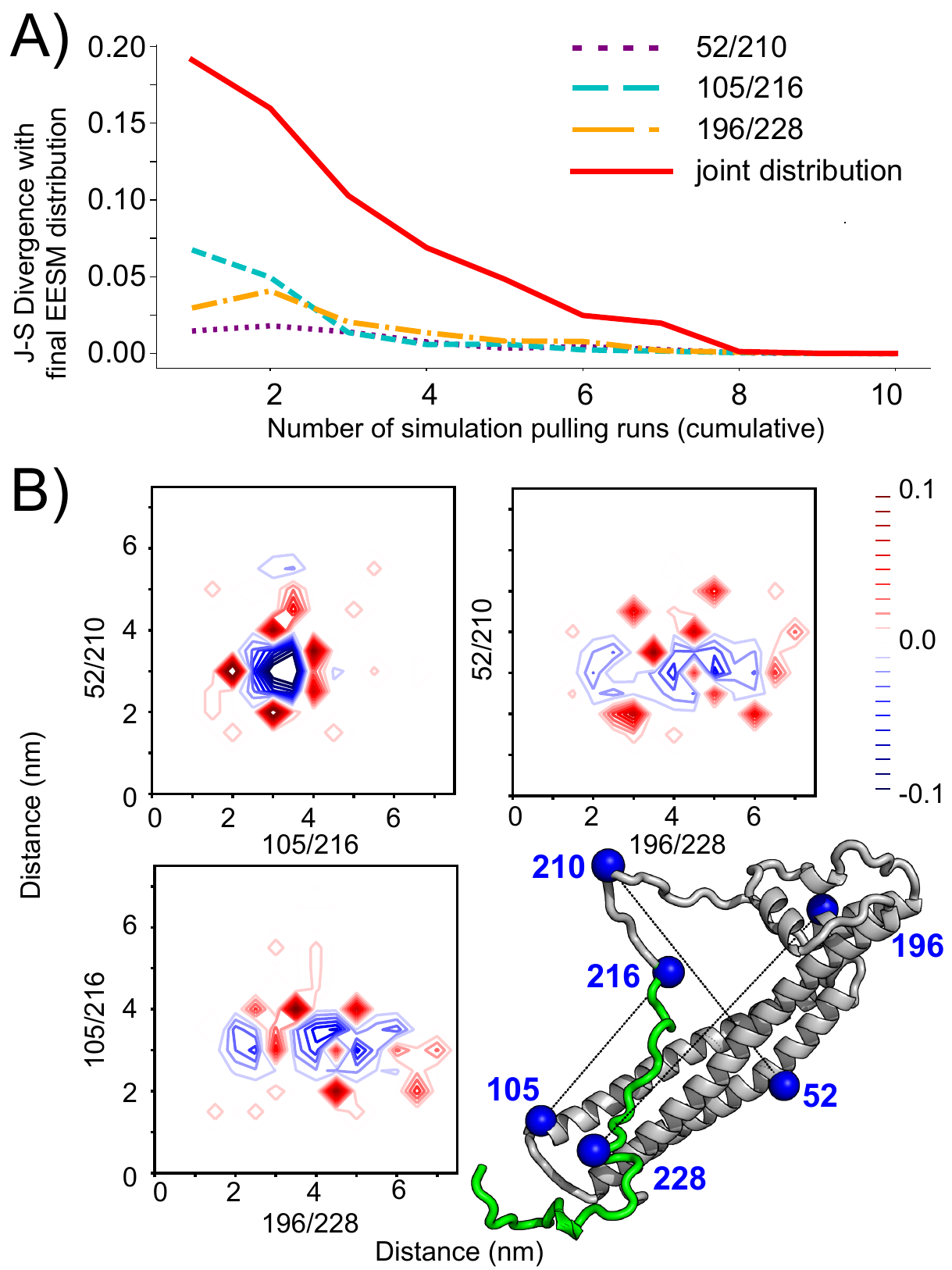}
    \caption{\textbf{EESM yields structural data on the syntaxin conformational ensemble.} EESM converges stably to a final estimate of the joint distribution of three experimentally-derived distributions (A). This final estimate differs substantially from the convolution of the experimental distributions, shown in (B) as contour plots of the difference $p_{\text{EESM}} - p_{\text{convolved }}$. The most down-weighted structure  is shown in the lower right. While this structure is allowed by the convolution of the experimental distributions, it is predicted to be low probability by EESM. Indeed, it is biochemically unlikely that syntaxin's terminal region (shown in green) would be highly disrupted while maintaining close contact with the other domains of the protein (shown in grey).}
    \label{fig:corr4}
\end{figure}

We performed an aggregate of 1.3 $\mu$s of molecular dynamics (MD) simulation and used EESM to estimate the joint probability distribution of the three DEER-derived distance distributions (projections of the joint space may be found in the Supplemental Material). EESM converges stably to a final estimate of the joint distribution over ten EESM iterations (Fig. \ref{fig:corr4}A). This final EESM estimate is quite different from the convolution of the experimental distributions (Fig. \ref{fig:corr4}B). By subtracting the EESM distribution from the experimental convolution, we identified sets of structures that are significantly down-weighted by the EESM method. These are structures predicted to have high probability under the assumption that the experimental distributions are independent, but have low probability after the EESM correction. The most down-weighted structure is shown in Fig. \ref{fig:corr4}B.  From inspection, this structure appears unlikely, as one of the major structural elements in syntaxin is disrupted. Specifically, we would not expect a low-energy conformation to have an unstructured backbone region that maintains contacts with the structured domain; instead, we expect complete dissociation of the end region (rendered in green in Fig. \ref{fig:corr4}B) while maintaining some secondary structure (see Supplement for further discussion). The presence or absence of such a structure can be tested via systematically-designed DEER experiments\cite{hays_refinement_2018} or other methods such as cross-linking mass spectrometry\cite{christie_2012}. Thus, EESM produces testable hypotheses about the syntaxin conformational ensemble.

We have developed a method, ensemble estimation from separate measurements, that can be used to infer the joint distribution of separately-acquired measurements and the conformational ensemble which optimally reproduces that distribution. The method was tested on a simplified model of an alternating-access transporter, where it efficiently yielded the true joint distribution and conformational ensemble \textit{a priori}. We found that EESM converged to the correct distribution within relatively few iterations (Fig \ref{fig:corr3}), confirming that EESM can be used to calculate the correlation structure of separately-measured distributions. 

EESM is particularly designed to estimate the conformational ensembles of systems where it is impossible to obtain a ground truth ensemble and joint distribution, as is the case for syntaxin-1a. We used EESM to estimate the syntaxin ensemble from three separately-acquired DEER distributions. We evaluated the method based on two criteria: its convergence behavior and its ability to generate testable hypotheses. We found that EESM converges smoothly to a final estimate of the joint distribution that is distinct from the convolved distributions (Fig. \ref{fig:corr4}A). Most importantly, the EESM-refined ensemble revealed structures that are predicted to have significantly lower probability in reality than one would have anticipated from the convolved distributions. These structures can be used immediately to design additional DEER experiments that would further refine the syntaxin ensemble.

Spectroscopic measurements that provide pair-wise distributions are a rich source of experimental data on heterogeneous ensembles. However, the utility of these measurements has been limited by the need to introduce and measure each label pair separately. EESM enables inference of the correlation structure between these separate measurements, greatly improving our ability to leverage such experiments to refine complex, flexible conformational ensembles.

\begin{acknowledgments}
The authors thank Z. Lu, E. Irrgang, K. Dubay, and D. Cafiso for helpful discussions.  This work was supported by NSF OAC-1835780 and NIH R01GM115790 to P.M.K, a Wallenberg Academy Fellowship to P.M.K., and a MolSSI Fellowship from the National Science Foundation (ACI-1547580) to J.M.H. Computational resources were provided by XSEDE TG-MCB150128.
\end{acknowledgments}




\bibliography{main}
\end{document}